\documentclass[a4paper,11pt]{article}
\usepackage{amsmath}
\usepackage{amssymb}
\usepackage{graphicx}
\usepackage{cite}

\newtheorem{thm}{Theorem}[section]

\newtheorem{prop}[thm]{Proposition}
\newtheorem{fact}[section]{Fact}
\newtheorem{thm*}[section]{Theorem}
\newtheorem{lem}[thm]{Lemma}

\newtheorem{con*}[section]{Conjecture}

\begin{document}
\title{A Reasonable Ab Initio Cosmological Constant Without Holography}
\author{Aaron Trout \\ \multicolumn{1}{p{.55\textwidth}}{\centering\emph{Department of Mathematics, \mbox{Chatham University}, \mbox{Pittsburgh PA, USA}}}}
\maketitle

\begin{abstract}
We give a well-motivated explanation for the origin of dark energy, claiming that it arises from a small residual negative scalar-curvature present even in empty spacetime. The vacuum has this residual curvature because spacetime is fundamentally discrete and there are more ways for a discrete geometry to have negative curvature than positive. We explicitly compute this effect in the well-known {\em dynamical triangulations} (DT) model for quantum gravity and the predicted cosmological constant $\Lambda$ agrees with observation.

We begin by almost completely characterizing the DT-model's vacuum energies in dimension three. Remarkably, the energy gap between states comes in increments of \[\Delta\mathcal{A} =\frac{\ell}{8\mathcal{V}}\] in natural units, where $\ell$ is the ``Planck length'' in the model and $\mathcal{V}$ is the volume of the universe. Then, using only vacua in the $N$ energy levels nearest zero, where $N$ is the universe's radius in units of $\ell$, we apply our model to the current co-moving spatial volume to get $|\Lambda| \approx 10^{-123}$.

This result comes with a rigorous proof and does not depend on any holographic principle or carefully tuned parameters. Our only unknown is the relative entropy of the low-energy states, which sets the sign of $\Lambda$. Numerical evidence strongly suggests that spacetime entropy in the DT-model is a decreasing function of scalar-curvature, so the model also predicts the correct sign for $\Lambda$.
\end{abstract}

\section{Introduction}
General relativity can be written in the Lagrangian formalism using the Einstein-Hilbert action, which in natural units is
\begin{equation}
	\label{EH_action}
	\mathcal{A_{E\!H}}(g) = \int_M \left[  \frac{1}{16\pi} \left( R - 2\Lambda \right) + \mathcal{L}_m \right]\! \mbox{dV}.
\end{equation}

\noindent Here $M$ is a closed $n$-manifold, $g$ a Lorentzian metric, $R$ scalar-curvature, $\Lambda$ the cosmological constant, and $\mathcal{L}_m$ the Lagrangian for matter. Note, both $R$ and $\mathcal{L}_m$ depend on $g$ while $\Lambda$ does not.

The critical points of $\mathcal{A_{E\!H}}$ are solutions to the field equations for general relativity. For the vacuum with $\Lambda =0$ we have action
\begin{equation}
	\label{EH_vacuum_action}
	\mathcal{A_{E\!H}^{\!\mbox{vac}}}(g) = \frac{1}{16\pi} \int_M \!R \,\mbox{dV}.
\end{equation}
\noindent Critical points of $\mathcal{A_{E\!H}^{\!\mbox{vac}}}$ satisfy:

\addtocounter{section}{-1}
\begin{fact} Any critical point $g$ of $\mathcal{A_{E\!H}^{\!\mbox{vac}}}$ is scalar-flat (R=0 everywhere) and $\mathcal{A_{E\!H}^{\!\mbox{vac}}}(g)=0$.
\label{classical_crit_value}
\end{fact}

In \cite{Regge1961} Regge gives a discrete version of $\mathcal{A_{E\!H}^{\!\mbox{vac}}}$ for piecewise-linear (PL) manifolds. We use the ``fully discrete'' version from the {\em dynamical triangulations} literature. Suppose $T$ is a combinatorial $n$-manifold homeomorphic to a fixed closed $n$-manifold $M$. We give $T$ a PL-metric by setting all edge-lengths to $\ell$, calling such a space a \textbf{triangulation} of $M$. In this model triangulations represent the possible \textbf{spacetime states}. Let $N_k(T)$ denote the number of $k$-simplices in $T$. Our action, which we call the \textbf{combinatorial Regge action} is
\begin{equation}
	\label{CR_action}
	\mathcal{A_{C\!R}}(T, \ell) = \frac{V_{n-2}(\ell)}{16\pi} \sum_{\tau^{n\!-\!2}\in T} \left( 2\pi - \theta_n \mbox{deg}(\tau^{n-2}) \right)
\end{equation}
\noindent where $V_k(\ell)$ is the volume of a regular $k$-simplex with side-length $\ell$, $\theta_n=\cos^{-1}(\frac{1}{n})$ is the {\em dihedral angle} in such a simplex, and $deg(\tau)$ is the number of $n$-simplices in $T$ with $\tau$ as a face. The success of the {\em Regge calculus} \cite{Rocek1981, Hamber94, Hamber95, Beirl97, Gentle02, Gentle12} and {\em (causal) dynamical triangulation} \cite{Agishtein92, Ambjorn92, Catterall94, Ambjorn04, Ambjorn06, Benedetti09, Ambjorn10, Ambjorn11} approaches to quantum gravity, both of which use Regge-type actions, gives us confidence that $\mathcal{A_{C\!R}}(T, \ell)$ corresponds to the classical notion of {\em total scalar-curvature}, at least at length-scales large compared with $\ell$.

\noindent 
\section{Action Spectrum}
\label{action_spectrum}

Henceforth, we work in dimension three. Let $\mathcal{T}(M)$ be the set of all triangulations of a closed 3-manifold $M$, and let $\mathcal{T}_{\!K}(M)$ denote those with exactly $K$ 3-simplices. We will write $\mathcal{A_{C\!R}}(T, \ell)$ in terms of the \textbf{mean edge-degree}
\begin{equation}\mu(T)=\frac{1}{N_1(T)}\sum_{e\in T} \mbox{deg}(e).\end{equation}
\noindent Some double-counting and algebra gives:
\begin{equation}
\mathcal{A_{C\!R}}(T,\ell) = \frac{3\ell}{4} N_3(T)\left( \frac{1}{\mu(T)} - \frac{1}{\mu^*}\right)
\label{NCR_mu_formula}
\end{equation} where $\mu^* = \frac{2\pi}{\theta_3}\approx 5.1$ is the \textbf{flat edge-degree}. This is the number of regular $3$-simplices needed around an edge to get a total dihedral angle of $2\pi$, the expected quantity in flat space. See Section \ref{spectrum_calcs} for complete details.

$\mathcal{A_{C\!R}}(T,\ell)$ is unbounded on $\mathcal{T}(M)$ so we define a volume-normalized version \begin{equation}\mathcal{A^{\!V\!N}_{C\!R}}(T, \ell) = \frac{\mathcal{A_{C\!R}}(T)}{\mbox{Vol}(T)} \label{normalized_action} \end{equation} where $\mbox{Vol}(T)=V_n(\ell)N_n(T)$ is the PL-volume of $T$. We will often write just $\mathcal{A^{\!V\!N}_{C\!R}}(T)$ or simply $\mathcal{A}_\mu$.\vspace{.1in}

\noindent {\sc Note:} Our normalization removes the dependence of $\mathcal{A_{C\!R}}$ on the {\em number} of 3-simplices at fixed $\mu$. The action still depends on $\ell$ as
\begin{equation}
	\label{NCR_scaling_equation}
	 \mathcal{A^{\!V\!N}_{C\!R}}(T, \ell) \propto \ell^{-2}.
\end{equation}

\noindent Because we wish to investigate effects resulting from the discreteness of spacetime, this dependence on $\ell$ is crucial. 

Using Euler-characteristic and double-counting arguments, for any $T\in\mathcal{T}_{\!K}(M)$ \begin{equation}N_0(T) = K\!\left( \frac{6}{\mu(T)} - 1 \right)\ , \ N_1(T) = K\! \frac{6}{\mu(T)}. \label{f_vector_K_mu}\end{equation} Details can be found in Section \ref{spectrum_calcs}. So, for fixed $K$, increasing $\mu$ decreases both $N_0$ and $N_1$. Thus, understanding the possible $\mathcal{A}_\mu$ means knowing which $N_0$ and $N_1$ occur in $\mathcal{T}(M)$.

In \cite{Walkup1970} Walkup defines ranges for $N_0$ and $N_1$ which must occur in $\mathcal{T}(M)$. These imply:
\addtocounter{section}{-1} \begin{thm*} Suppose $\mathcal{A}_6 < x < \mathcal{A}_{4.5}$. For all large enough $K$ there are triangulations $T^+$ and $T^-$ in $\mathcal{T}_{\!K}(M)$ with \begin{equation}\mu(T^+)=\frac{6K}{N_1(T^+)}\end{equation} and \begin{equation}\mu(T^-)=\frac{6K}{N_1(T^-)} = \frac{6K}{N_1(T^+)-1}\end{equation} and for which $\mathcal{A}_{\mu(T^-)} \leq x \leq \mathcal{A}_{\mu(T^+)}$. 
\label{adjacent_action_properties}
\end{thm*}
See Section \ref{spectrum_calcs} for a detailed proof.

Let $\mathcal{N}^-_{\!K,x}$ and $\mathcal{N}^+_{\!K,x}$ denote the number of states in $\mathcal{T}_{\!K}(M)$ with the same action as  $T^-$ and $T^+$ respectively. By equation (\ref{f_vector_K_mu}), $\mu(T)=\frac{6K}{N_1(T)}$ so $T^+$ and $T^-$ have actions as close as possible to $x$ at fixed $K$. Also, note that in natural units $\mathcal{A}_6\approx -0.19$ and $\mathcal{A}_{4.5}\approx 0.17$ so the endpoints represent {\em enormous} energy densities. The energy gap $\Delta\mathcal{A} = \mathcal{A}_{\mu(T^+)} - \mathcal{A}_{\mu(T^-)}$ is given by
\begin{equation}
\Delta\mathcal{A} = \frac{3\ell}{4 V_3(\ell)} \left( \frac{1}{6K} \right) = \frac{3\sqrt{2}}{4}\frac{1}{\ell^2K}.
\label{delta_A_KL_eqn}
\end{equation}
\noindent Since $V_3(\ell)K = \mbox{Vol}(T)$ this is also
\begin{equation}
\Delta\mathcal{A} = \frac{1}{8}\frac{\ell}{\mbox{Vol}(T)}.
\label{delta_A_Lvol_eqn}
\end{equation}

\section{Degeneracy Data}
\begin{figure*}
\begin{center}
\includegraphics[angle=-90, width=\textwidth]{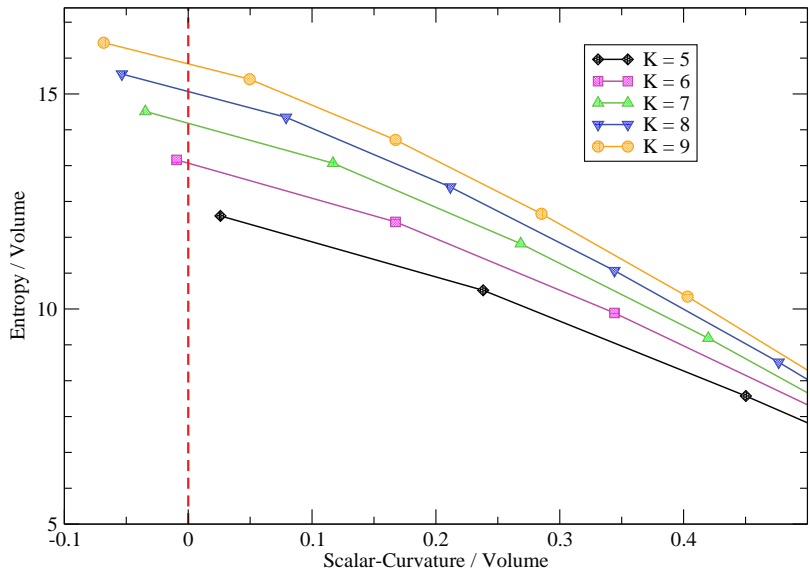}
\end{center}
\caption{Spacetime entropy per volume $S=\frac{1}{V_3K}\ln\left| \mathcal{T}_{\!K}(S^3)\right|$ for $S^3$ at various $K$, plotted versus scalar-curvature per volume $\mathcal{A}_\mu$ at $\ell=1$. Data comes from a census created by Benjamin Burton \cite{Burton2004, Burton2011}. Many thanks to Henry Segerman for providing the underlying data, extracted from the ($\sim 47$ million) triangulations of $S^3$ with at most 9 tetrahedra. {\sc Note}: Burton's definition of triangulation is less restrictive than ours, requiring only that $T$ be an abstract simplicial complex homeomorphic to $M$. We believe this is an unimportant technical issue and any census using our definition would show the same dependence of entropy on curvature.}
\label{entropy_vs_curvature}
\end{figure*}

To compute the expected value of an observable over some set of energy states, we need to know the relative degeneracies of those states.  Data from censuses of 3-manifold triangulations \cite{Burton2004, Burton2011} lead us to conjecture:
\addtocounter{section}{-1} \begin{con*} For any $\mathcal{A}_6 < x < \mathcal{A}_{4.5}$ the limit \begin{equation}\lim_{K\rightarrow \infty} \frac{\mathcal{N}^+_{K,x}}{\mathcal{N}^-_{K,x}} = C(x)\end{equation}
exits. Moreover, $C$ is a continuous function of $x$ with $C(x)<1$. 
\label{degeneracy_con}
\end{con*}
That is, we conjecture that the degeneracies decrease roughly exponentially near each action $x$. This effect can be seen in Figure \ref{entropy_vs_curvature} which plots the (per-volume) spacetime entropy
\begin{equation}
	S = \frac{1}{V_3(\ell)K}\ln\left| \mathcal{T}_{\!K}(M)\right|
\end{equation} as a function of the (per-volume) scalar-curvature $\mathcal{A}_\mu$ for $M=S^3$ at various $K$ and $\ell=1$. Notice the downward slope of the curves.

\section{The Nearly-Flat Model}
Often in quantum gravity it is difficult to write down a well-behaved (or even well-defined) partition function. Here, there is no such problem. Inspired by Fact \ref{classical_crit_value}, we will create a system {\em obviously} dominated by states with $\mathcal{A}_\mu \approx \mathcal{A}_{\mu^*} = 0$, the ``correct'' classical value. We know the action values, so the only question is how many energy states $N$ to use on either side of zero. Our $N$ should be a large dimensionless number involving only Vol$(T)$ and $\ell$. A natural choice is
\begin{equation} \label{num_states} N=\frac{\mbox{Vol}(T)^{1/3}}{\ell}. \end{equation} This is roughly the {\em radius of the universe} in units of $\ell$. If Conjecture \ref{degeneracy_con} holds and $N$ is large we have
\begin{equation}
\label{expected_action} \langle \mathcal{A}_\mu \rangle \approx -N\Delta\mathcal{A}.
\end{equation} 
See Section \ref{NF_calcs} for the complete calculation. Note, unless $C(x)$ is {\em very} close to one near $x=0$, only the sign of $\langle \mathcal{A}_\mu \rangle$ comes from our conjecture. Also, in the current co-moving volume $N\gg 1$ and $\langle \mathcal{A}_\mu \rangle \approx 0$ so our model contains a large number of nearly ``classical'' states.

\subsection{The Cosmological Constant}

Let us discuss the physical meaning of $\langle \mathcal{A}_\mu\rangle$. If $\mathcal{A_{C\!R}}$ measures total scalar-curvature then $ \mathcal{A}_\mu$ is average scalar-curvature per volume. $\mathcal{A}_\mu$ contains no built-in $\Lambda$ and we force it near zero in an unbiased way. By Fact \ref{classical_crit_value} one would expect $\langle \mathcal{A}_\mu\rangle=0$ for our model, but as we have seen this does not occur. Moreover, the failure results from the relative entropy of {\em action values} rather than the detailed dynamics of the ``metric'' $T$. Since everything in the classical action $\mathcal{A_{E\!H}}$ except $\Lambda$ depends on the metric $g$, the structure of $\mathcal{A_{E\!H}}$ practically {\em demands} we interpret $\langle \mathcal{A}_\mu\rangle$ as an emergent $\Lambda$ given by
\begin{equation}
\langle \mathcal{A}_\mu\rangle = -2\Lambda.
\label{expected_action_lambda_relation}
\end{equation}
Now, we take $\ell \approx 1.6 \times 10^{-35}\mbox{\,m}$ to be Planck's length and $\mbox{Vol}(T) \approx 3.5 \times 10^{80} \mbox{\,m}^3$ to be the universe's co-moving spatial volume. This gives, by equations (\ref{delta_A_Lvol_eqn}) and (\ref{num_states}) -- (\ref{expected_action_lambda_relation}) \begin{equation}\Lambda \approx 10^{-123}\end{equation} in agreement with observation.
\section{Discussion}

It is emphatically {\em not} the purpose of this paper to advocate ``triangulations'' as the ultimate structure of spacetime. Indeed, the effect we describe could well occur in other discrete-spacetime theories, such as {\em loop quantum gravity}, {\em spin foam} models, and {\em causal dynamical triangulations} to name a few.

Our calculation of $\Lambda$ has some interesting features which we now discuss. First, we cheated by using the {\em current} value of the universe's co-moving {\em spatial} volume. Does this mean $\Lambda$ actually changes over time? The fundamental dimensionless parameter here is \begin{equation}\alpha_G = \mbox{Vol}(T)^{2/3}\Lambda\end{equation} which we dub the \textbf{geometric fine-structure constant} since it controls the scale of entropic perturbations from scalar-flattness caused by the discreteness of spacetime. If our use of the co-moving volume is acceptable, then our model predicts
\begin{equation}
\Lambda(t)\propto \mathcal{V}(t)^{-2/3}
\end{equation}
where $\mathcal{V}(t)$ is the co-moving volume at a co-moving observer's proper time $t$. A time-varying $\Lambda$ seems strange, but that does not make it false. Indeed, perhaps discrete spacetime effects caused the rapid expansion of the early universe postulated in inflationary big-bang theories.

Second, like other recent work \cite{Cohen99, Horvat04, Li10, Easson11, Castorina12} on $\Lambda$, our model involves both {\em global} and {\em local} properties of the universe. Many of these models also suggest that $\Lambda$ varies over time. These other approaches are quite different in detail from ours, but the broad similarities are striking. Perhaps we are all somehow pointing at the same underlying issue. 

Third, thinking of gravity as an emergent and/or entropic force is very popular at the moment \cite{Steinacker07, Hu09, Steinacker09, RongGen10, Gao10, Nicolini10, Verlinde11}. Our model provides a concrete example of treating general relativity as a {\em mean-field approximation} (in this case, to the underlying dynamic-triangulation style geometry). However, we emphasize that our approach is significantly less ambitious than Verlinde's {\em entropic gravity} program \cite{Verlinde11} and does not rely on any assumed holographic principle.

Finally, there are many loose ends to our story. We would certainly prefer 4-dimensional results, and the sign of $\Lambda$ depends on Conjecture \ref{degeneracy_con} making its proof an important goal. There is also the issue of our choice for the number of nearly-flat states $N$. Taking this to be the radius of the universe in natural units seems similar to the IR cutoff imposed on quantum field theory in many explanations of $\Lambda$. What is the connection between these two techniques, and can our choice for $N$ be justified using appropriate detailed dynamics on $\mathcal{T}(M)$? There remains much work to be done.

\section{Action Spectrum Proofs}
\label{spectrum_calcs}

In this section we give detailed proofs for our characterization of the energy spectrum for the dynamical-triangulations model in dimension three. We begin with a little double-couting result:

\begin{lem} If $T\in \mathcal{T}(M)$ we have

\begin{equation}
	\label{mu_formulas}
	\mu(T) =\frac{6N_3(T)}{N_1(T)} = \frac{3N_2(T)}{N_1(T)}.
\end{equation}
\end{lem}

\noindent {\sc Proof}: Suppose we examine each edge in $T$, placing a check-mark on the tetrahedra around it. Clearly, we have made $\sum_e \mbox{deg}(e)$ marks. However, since each tetrahedra has six edges, it is marked six times and we must also have $6N_3(T)$ marks. Dividing by $N_1(T)$ gives our first equation. Now, for each triangle imagine placing marks on the two tetrahedra meeting at that triangle. We have obviously placed $2N_2$ marks. Since each tetrahedra has four triangular faces we have also made $4N_3$ marks. So, $N_2=2N_3$ giving us the second equality. $\Box$

Using the first part of this equation we can express $\mathcal{A_{C\!R}}(T)$ as a function of our preferred variables: number of $3$-simplices and mean edge-degree.
\begin{prop}
\label{CR_mu_relation}
The combinatorial Regge action is related to the mean edge-degree according to \begin{equation}
\mathcal{A_{C\!R}}(T,\ell) = \frac{3\ell}{4} N_3(T)\left( \frac{1}{\mu(T)} - \frac{1}{\mu^*}\right)
\label{NCR_mu_formula}
\end{equation} where $\mu^* = \frac{2\pi}{\theta_n}$.
\end{prop}

\noindent {\sc Proof}: We begin by writing down the combinatorial Regge action $\mathcal{A_{C\!R}}$ in dimension three.\begin{equation}
	\label{CR_action}
	\mathcal{A_{C\!R}}(T, \ell) = \frac{\ell}{16\pi} \sum_{e\in T} \left( 2\pi - \theta_3 \mbox{deg}(e) \right).
\end{equation}
\noindent Distributing the sum into the summand we get 
\begin{equation}
	\mathcal{A_{C\!R}}(T, \ell) = \frac{\ell}{16\pi} \left( 2\pi N_1(T) - \theta_3 \sum_{e\in T} \mbox{deg}(e) \right).
\end{equation}
\noindent Now, by equation (\ref{mu_formulas}) we can replace $N_1(T)$ with $\frac{6N_3(T)}{\mu(T)}$ and the remaining summation by $6N_3(T)$. This gives
\begin{equation}
	\mathcal{A_{C\!R}}(T, \ell) = \frac{\ell}{16\pi} \left( 2\pi\frac{6N_3(T)}{\mu(T)} - \theta_3 6N_3(T) \right).
\end{equation}
Pulling out a factor of $12\pi N_3(T)$ gives
\begin{equation}
	\mathcal{A_{C\!R}}(T, \ell) = \frac{3\ell}{4}N_3(T) \left( \frac{1}{\mu(T)} - \frac{\theta_3}{2\pi} \right)
\end{equation}
which is just
\begin{equation}
	\mathcal{A_{C\!R}}(T, \ell) = \frac{3\ell}{4}N_3(T) \left( \frac{1}{\mu(T)} - \frac{1}{\mu^*} \right)
\end{equation}
as desired. $\Box$

Next we write the number of vertices $N_0(T)$ and edges $N_1(T)$ as a functions of $N_3(T)$ and $\mu(T)$.
\begin{lem} If $T\in \mathcal{T}(M)$ we have \[N_0(T) = N_3(T)\left( \frac{6}{\mu(T)} - 1 \right) \ \ \mbox{and}\ \ \ N_1(T) = N_3(T) \frac{6}{\mu(T)}.\] 
\label{f_vector_lem}
\end{lem}
\noindent {\sc Proof}: Any 3-manifold has Euler-characteristic zero. This means
\begin{equation}
\label{euler_char}
N_0(T) - N_1(T) + N_2(T) - N_3(T)=0.
\end{equation}
Now, we use equation (\ref{mu_formulas}) to replace $N_2(T)$ by $2N_3(T)$ to get
\begin{equation}
\label{N013_formula}
N_0(T) - N_1(T)+ N_3(T)=0.
\end{equation}
Using equation (\ref{mu_formulas}) again to replace $N_1(T)$ by $\frac{6N_3(T)}{\mu(T)}$ gives
\begin{equation}
N_0(T) - \frac{6N_3(T)}{\mu(T)} + N_3(T)=0
\end{equation}
which can be rearranged to produce
\begin{equation}
N_0(T) = N_3(T)\left( \frac{6}{\mu(T)} - 1 \right)
\end{equation}
as desired. Next, we plug this formula for $N_0(T)$ into equation (\ref{N013_formula}) to get
\begin{equation}
N_3(T)\left( \frac{6}{\mu(T)} - 1 \right) - N_1(T)+ N_3(T)=0
\end{equation}
which simplifies to
\begin{equation}
N_1(T) = N_3(T) \frac{6}{\mu(T)}
\end{equation}
completing the proof. $\Box$

Now we are ready for the crucial step. If we wish to characterize the action values we need to know that triangulations {\em actually exist} in $\mathcal{T}(M)$ with particular $N_0$ and $N_1$. This task is handled by a famous 1970 result by Walkup.

\begin{thm}[Walkup] For every closed 3-manifold $M$ there is a smallest integer $\gamma^*(M)$ so that any two positive integers $N_0$ and $N_1$ which satisfy \[{N_0 \choose 2} \geq N_1 \geq 4N_0 + \gamma^*(M) \] are given by $N_1 = N_1(T)$ and  $N_2 = N_2(T)$ for some $T \in \mathcal{T}(M)$. The quantity $\gamma^*(M)$ is a topological invariant which satisfies $\gamma^*(M)\geq -10$ for all closed 3-manifolds $M$.
\label{walkup_thm}
\end{thm}

\noindent  Note that $\gamma^*(M)$ is known for many manifolds $M$, although we will not need this information. Using Walkup's result we can prove:

\begin{lem} Let $M$ be a closed 3-manifold and $K$ a fixed positive integer. For each integer $N_1$ which satisfies \[K+\frac{1}{2}\left( 3 + \sqrt{9+8K} \right) \leq N_1 \leq \frac{1}{3}\left( 4K - \gamma^*(M) \right) \] there is some triangulation $T \in \mathcal{T}_{\!K}(M)$ with $N_1 = N_1(T)$.
\label{N1_range_lem}
\end{lem}

\noindent {\sc Proof}: Suppose that $N_1 \leq \frac{1}{3}\left( 4K - \gamma^*(M) \right)$ and define $N_0 = N_1 - K$. A bit of simple algebra tells us
\begin{equation}
\label{N1_lower_bound}
N_1\geq 4N_0 + \gamma^*(M).
\end{equation}
Now, consider the upward opening parabola \[f(m) = {m \choose 2}-m-K\] which has largest root \[ m_0 = \frac{1}{2}\left( 3 + \sqrt{9 + 8K} \right). \]
Our hypothesis that $N_1 \geq K+\frac{1}{2}\left( 3 + \sqrt{9+8K} \right)$ implies $N_0 \geq \frac{1}{2}\left( 3 + \sqrt{9+8K} \right)$, so that $N_0 \geq m_0$.  Since $m_0$ is the largest root of an upward opening parabola, we conclude $f(N_0)\geq 0$. By our definition of $f$ and $N_0$, this tells us
\begin{equation}
\label{N1_upper_bound}
{N_0 \choose 2} \geq N_1.
\end{equation}
By Walkup's theorem, the inequalities (\ref{N1_lower_bound}) and (\ref{N1_upper_bound}) imply that some $T\in \mathcal{T}(M)$ has $N_0 = N_0(T)$ and $N_1 = N_1(T)$. Finally, by Lemma \ref{f_vector_lem}, we have $N_3(T) = N_1(T) - N_0(T) = K$ so that $T$ is in $\mathcal{T}_{\!K}(M)$ as desired. $\Box$\vspace{.15in}

\noindent We can use Lemma \ref{N1_range_lem} to show that for large enough $K$ there are triangulations in $\mathcal{T}_{\!K}(M)$ with mean edge-degree just on either side of any value in the interval $(4.5, 6)$.

\begin{lem} Fix any real number $4.5 < m < 6$. For all sufficiently large $K$ there are triangulations $T_1$ and $T_2$ in $\mathcal{T}_{\!K}$ with \begin{equation}\mu(T_1)=\frac{6K}{N_1(T_1)}\end{equation} and \begin{equation}\mu(T_2)=\frac{6K}{N_1(T_2)} = \frac{6K}{N_1(T_1)-1}\end{equation} and which satisfy $\mu(T_1) \leq m \leq \mu(T_2)$.
\end{lem}

\noindent {\sc Proof}: We begin with the bound given in Lemma \ref{N1_range_lem} \[K+\frac{1}{2}\left( 3 + \sqrt{9+8K} \right) \leq N_1 \leq \frac{1}{3}\left( 4K - \gamma^*(M) \right). \] We know that any $N_1$ in this range is $N_1(T)$ for some $T\in \mathcal{T}_K(M)$. Now, dividing by $6K$ and taking reciprocals gives an equivalent set of inequalities \[\frac{6K}{\frac{1}{3}\left( 4K  - \gamma^*(M) \right)} \leq \frac{6K}{N_1} \leq \frac{6K}{K+\frac{1}{2}\left( 3 + \sqrt{9+8K} \right)}.\] By equation (\ref{mu_formulas}) the quantity in the middle is just the mean edge-degree $\mu(T)$.  As $K\rightarrow\infty$, the LHS converges to $4.5$ and the RHS to $6$, so for sufficiently large $K$ we know triangulations exists with $\mu$ on either side of $m$. Finally, for fixed $K$, $\mu(T)$ is a decreasing function of $N_1(T)$ so the $\mu(T_1)$ and $\mu(T_2)$ values must be of the stated form. $\Box$\vspace{.1in} 

\noindent This tells us we can find triangulations with actions which ``bracket'', as closely as possible, any number in the interval $\left( \mathcal{A}_6, \mathcal{A}_{4.5}\right)$.

\addtocounter{section}{-5}
\begin{thm*} Suppose $\mathcal{A}_6 < x < \mathcal{A}_{4.5}$. For all large enough $K$ there are $T_1$ and $T_2$ in $\mathcal{T}_{\!K}(M)$ with \begin{equation}\mu(T_1)=\frac{6K}{N_1(T_1)}\end{equation} and \begin{equation}\mu(T_2)=\frac{6K}{N_1
(T_2)} = \frac{6K}{N_1(T_1)-1}\end{equation} and which satisfy $\mathcal{A}_{\mu(T_2)} \leq x \leq \mathcal{A}_{\mu(T_1)}$. 
\label{adjacent_action_properties}
\end{thm*}
\addtocounter{section}{4}

\section{Calculations for the Nearly-Flat Model}
\label{NF_calcs}
In this section we give detailed calculations for the nearly-flat model. 
\begin{lem} If $0<r<1$ and $N\gg 1$ then
\begin{equation}
r^N \sum_{n=-N}^{N}nr^n \approx -N\frac{1}{1-r}
\end{equation}
\end{lem}
\noindent {\sc Proof}: We calculate
\begin{align*}
r^N \sum_{n=-N}^{N}nr^n & = r^{N+1} \sum_{n=-N}^{N}nr^{n-1} \\
& = r^{N+1} \frac{d}{dr}\sum_{n=-N}^{N}r^n \\
& \approx r^{N+1} \frac{d}{dr} \left( \frac{r^{-N}}{1-r}\right) \\
& = r^{N+1} \left( \frac{-N(1-r)r^{-N-1} + r^{-N}}{1-r}\right) \\
& = \frac{-N(1-r) + r}{1-r} \\
& \approx -\frac{N}{1-r}.
\end{align*}
This completes the proof. $\Box$ \vspace{.1in}

\noindent Now we can calulate the expected action, assuming our conjecture, which we restate for the reader's convenience.

\addtocounter{section}{-5}
\begin{con*} For any $\mathcal{A}_6 < x < \mathcal{A}_{4.5}$ the limit \begin{equation}\lim_{K\rightarrow \infty} \frac{\mathcal{N}^+_{K,x}}{\mathcal{N}^-_{K,x}} = C(x)\end{equation}
exits. Moreover, $C$ is continuous with $C(x)<1$. 
\label{degeneracy_con}
\end{con*}
Using this conjecture, we can show the following.
\addtocounter{section}{4}

\begin{lem} If our conjecture is true, the expected action in the $N$ state nearly-flat model when $N\Delta\mathcal{A}\approx 0$ and $N \gg 1$ is
\begin{equation}
\langle \mathcal{A_\mu} \rangle \approx - N\Delta\mathcal{A}
\end{equation}
\end{lem}
\noindent {\sc Proof}: We write the standard formula for the expected value
\begin{equation}
\langle \mathcal{A}_\mu \rangle = \frac{\sum_{n=-N}^{N}\alpha_n\mathcal{N}_i e^{i \alpha_n}}{\sum_{n=-N}^{N}\mathcal{N}_i e^{i \alpha_i}}
\end{equation}
where $\alpha_k = k \Delta\mathcal{A}$ are the action values and $\mathcal{N}_k$ their degeneracies. Since we assumed $N\Delta\mathcal{A} \approx 0$ each $\alpha_k\approx 0$ and we have 
\begin{equation}
\langle \mathcal{A}_\mu \rangle \approx \frac{\sum_{n=-N}^{N}\alpha_n\mathcal{N}_i }{\sum_{n=-N}^{N}\mathcal{N}_i }.
\end{equation}
Using $N\Delta\mathcal{A} \approx 0$ and the assumed continuity of $C(x)$ gives
\begin{equation}
\langle \mathcal{A}_\mu \rangle \approx \frac{\sum_{n=-N}^{N}\alpha_n C^n }{\sum_{n=-N}^{N}C^n }
\end{equation}
where $C=C(0)<1$ and we have divided the top and bottom by $\mathcal{N}_0$. Since $N\gg 1$ we use the sum of a geometric series to get
\begin{equation}
\langle \mathcal{A}_\mu \rangle \approx \frac{C^N}{1-C}\sum_{n=-N}^{N}\alpha_n C^n.
\end{equation}
Finally, we apply the previous lemma to get the desired result:
\begin{equation}
\langle \mathcal{A_\mu} \rangle \approx - N\Delta\mathcal{A}
\end{equation}
completing the proof. $\Box$

\bibliography{dark_energy}
\bibliographystyle{plain}

\end{document}